\documentclass[12pt]{article}
\usepackage{amsfonts,amsmath,amssymb}
\usepackage{scalerel}[2016/12/29]
\usepackage{graphicx}
\usepackage{tikz}
\usepackage{braket}

\usetikzlibrary{arrows}
\usetikzlibrary{decorations.markings}

\numberwithin{equation}{section}

\def\be{\begin{equation}}
\def\ee{\end{equation}}
\def\bea{\begin{eqnarray}}
\def\eea{\end{eqnarray}}

\def\half{{\textstyle {1\over 2}}}

\def\bb{\hskip -0.5mm}

\usepackage{amsmath}
\usepackage{epsfig,graphicx}
\usepackage{graphicx}
\usepackage{tikz}

\def\be{\begin{equation}}
\def\ee{\end{equation}}
\def\bea{\begin{eqnarray}}
\def\eea{\end{eqnarray}}

\def\q{\mathrm{Q}}
\def\z{\mathrm{Z}}

\def\tF{\hat{F}}
\def\tG{\hat{G}}
\def\ttG{\hat{\hat{G}}}

\def\tH{\hat{H}}

\begin{document}

\thispagestyle{plain}

\title{\bf\Large Hamiltonian and exclusion statistics approach to discrete forward-moving paths}

\author{St\'ephane Ouvry$^*$ \  {\scaleobj{0.9}{\rm and}} Alexios P. Polychronakos$^\dagger$}

\maketitle

\begin{abstract}
We use a Hamiltonian (transition matrix) description of height-restricted Dyck paths on the plane in which generating
functions for the
paths arise as matrix elements of the propagator to evaluate the length and area generating function for paths with arbitrary
starting and ending points, expressing it as a rational combination of determinants. Exploiting a connection between random walks and
quantum exclusion statistics that we previously established, we express this generating function in terms of grand partition functions
for exclusion particles in a finite harmonic spectrum and present an alternative, simpler form for its logarithm that makes its polynomial structure explicit.
\end{abstract}

\noindent
* LPTMS, CNRS,  Universit\'e Paris-Sud, Universit\'e Paris-Saclay,\\ \indent 91405 Orsay Cedex, France; {\it stephane.ouvry@u-psud.fr}

\noindent
$\dagger$ Department of Physics, the City College \& The Graduate Center of CUNY,\\ \indent New York,
NY 10031, USA; 
{\it apolychronakos@ccny.cuny.edu}
\vskip 1cm

\vfill
\eject

\section{Introduction}

Random walks of given length and area on planar lattices are of inherent mathematical and physical interest. In mathematics, their statistics and generating functions are the subject of intense study. In physics, they arise either in actual diffusion processes or in the quantum mechanics of particles moving in a periodic two-dimensional potential. The
Hofstadter problem is the canonical example of the latter, leading to the famous ``butterfly'' energy spectrum \cite{Hofstadter}.

In physical contexts, random walks are generated through the action of a Hamiltonian on the Hilbert space of the
system. This connection was used to study the enumeration of closed walks of given length and (algebraic) area
on the square lattice. Such walks are generated by the Hofstadter Hamiltonian, and their properties can be derived
from the study of the secular determinant of the Hamiltonian. The area enumeration generating function
for walks of given length was derived in \cite{nous} in terms of a set of factors (the so-called Kreft coefficients \cite{Kreft})
extracted from the secular determinant, leading to explicit albeit complicated expressions.

A remarkable connection was made in \cite{emeis} between a general class of two-dimensional walks and quantum
mechanical particles obeying generalized {\it exclusion statistics} with exclusion parameter $g$ ($g\bb=\bb 0$ for bosons, $g\bb=\bb 1$ for fermions, and higher $g$ means a stronger exclusion beyond Fermi) depending on the type of walks.
Exclusion statistics was proposed by Haldane \cite{Haldane} as a distillation of the statistical mechanical properties of Calogero-like spin systems. The relevance of generalized quantum statistics to Calogero particles was first pointed out in \cite{NRBos}. Exclusion statistics also emerges in the context of anyons projected on the lowest Landau level of a strong magnetic field \cite{Das}, and has been extended to more general systems \cite{Poly}. Remarkably, the algebraic area considerations of a class
of lattice walks directly map to the statistical mechanics of particular many-body systems with exclusion statistics. 

In the present work we make use of both the Hamiltonian description of random walks and the exclusion statistics
connection to study the generating function of a family of walks usually referred to as Dyck paths and their height-restricted
generalizations 
% \cite{Stan,BM,BIP,OwPr,Bach,Nech,KP}
$\cite{Stan}{-}\cite{KP}$\footnote{The literature on Dyck and related Lukasiewicz paths is quite extensive. For a comprehensive list of papers we refer
the reader to T. Prellberg's site http://www.maths.qmul.ac.uk/{\tiny$\sim$}tp/}.
These are walks that propagate one step in
the horizontal direction (``time'') and one step either up or down in the vertical direction (``height'') but without dipping below
a ``floor'' at height zero nor exceeding a ``ceiling'' of maximal height. Paths that start and end at the floor are usually termed
``excursions'', while more general paths are ``meanders''.
The Hamiltonian method for forward-moving paths is equivalent to the transition matrix formulation, which has been used in
previous work to calculate the length generating function for such walks. In the present work we extend these results
to length {\it and} area generating functions for meanders with arbitrary starting and ending points.
Although the results for the length generating function for Dyck meanders, and the length and area generating function
for Dyck excursions have been obtained before, our expression of the length and area generating function for arbitrary
meanders is, to the best of our knowledge, a new result.

Further, using the connection to exclusion statistics that we established earlier allows us to express the generating functions
in terms of
statistical mechanical properties of known and relatively simple systems, and to derive alternative forms for their logarithms in terms of sums over compositions (i.e., ordered partitions) of the integer length of the path that make explicit their polynomial structure.
Again, to the best of our knowledge, this connection and corresponding expressions are original.

In the next section we set up the Hamiltonian description of walks and establish the fundamental expression of their
generating functions in terms of matrix elements of the propagator, while in section 3 we derive the basic determinant
formula for the generating functions and examine several special cases. In section 4 we review the exclusion statistics
connection and use it to express the basic building block of the generating functions, the secular determinant, in terms of
grand partition functions, while in section 5 we apply the method to derive explicit expressions for the generating functions
and their logarithms.
In section 6 we complete the analysis by extending it to generating functions counting also the ``touchdowns'' of the paths.
We conclude with some remarks on a connection of the walks in the present work with the group theoretical problem of
the composition of a large number of spins and $q$-extensions of Schur functions, and present some directions for future work.

\section{Walk generating Hamiltonian}

Consider a Hamiltonian acting on a $k+1$-dimensional Hilbert space of the form
\be
H_k = V_k^\dagger U_k + U_k V_k
\ee
where
\be
U_k = \begin{pmatrix}
1&\;\; 0 &\;\; 0 & \cdots & 0 & 0 \\
0 &\;\; \q  &\;\; 0& \cdots & 0 & 0 \\
0 &\;\; 0 &\;\; \q^2  & \cdots & 0 & 0 \\
\vdots &\;\; \vdots &\;\; \vdots & \ddots & \vdots & \vdots \\
0 &\;\; 0 &\;\; 0  & \cdots &\q^{k-1}  & 0 \\
0 &\;\; 0 &\;\; 0  & \cdots & 0 & \q^{k}\\
\end{pmatrix}~,~~~
\label{umat} V_k =  \begin{pmatrix}
0\;\;& 1\;\, & 0\;\; & 0\cdots\;\; & 0\;\; & 0\; \\
0 \;\;& 0 \;\;&1\;\;& 0\cdots \;\;& 0 \;\;& 0\; \\
0 \;\;& 0\;\; & 0\;\; &1 \cdots\;\; & 0 \;\;& 0\; \\
\vdots \;\;& \vdots \;\;& \vdots \;\;& \ddots \;\;& \vdots\;\; & \vdots\; \\
0\;\; & 0 \;\;& 0 \;\; &0 \cdots\;\; & 0\;\; & 1 \;\\
0\;\;& 0 \;\;& 0 \;\; &0 \cdots \;\;& 0\;\;& 0\; \\
\end{pmatrix}
\ee
$U_k$ and $V_k$ are similar to the usual ``clock'' and ``shift'' matrices of the ``quantum torus'' algebra, with the
differences that the parameter $\q$ need not be a root of unity, and $V_k$ is missing the lower left element that would make
it cyclic and invertible. They obey the relations
\be
V_k U_k = \q\, U_k V_k ~,~~ V_k V_k^\dagger = 1 - \ket{k}\bb \bra{k} ~,~~ V_k^\dagger V_k = 1 - \ket{0}\bb \bra{0}
\ee
where $\ket{n}\bb\bra{n}$ is the projector on the $n$-th element of the space ($n=0,1,\dots,k$). 
The Hamiltonian has the explicit off-diagonal form
\be
{H}_k = 
\begin{pmatrix}
0\;\;& 1\;\, & 0\;\; & 0\;\; \cdots\;\; & 0\;\; & 0\; \\
1 \;\;& 0 \;\;&\q\;\;& 0\;\; \cdots \;\;& 0 \;\;& 0\; \\
0 \;\;& \q\;\; & 0\;\; &\q^{2} \;\cdots\;\; & 0 \;\;& 0\; \\
\vdots \;\;& \vdots \;\;& \vdots \;\;& \;\;\ddots & \vdots\;\; & \vdots\; \\
0\;\; & 0 \;\;& 0 \;\; &0\;\;  \cdots& 0\;\; & \q^{k-1} \;\\
0\;\;& 0 \;\;& 0 \;\; &0 \;\;\cdots & \q^{k-1}\;\;& 0\; \\
\end{pmatrix}
\label{hamk}\ee
and is symmetric. It is also Hermitian for real $\q$, although this is of no import for our considerations.

{\begin{figure}
\centering 
{\hskip -2cm
\begin{tikzpicture}[scale=1]

% square grid
\draw[help lines, gray] (0,-0.02) grid (13.5,4.01);

% axes i, j and k
% position: (below/above) + left/right
\tikzset{big arrow/.style={decoration={markings,mark=at position 1 with {\arrow[scale=3,#1,>=stealth]{>}}},postaction={decorate},},big arrow/.default=black}
% Different styles for arrows: >=to (default), stealth, latex
% Alternative type: \tikzset{big arrow/.style={decoration={markings,mark=at position 1 with {\arrow[scale=2,#1,>=to]{>}}},postaction={decorate},},big arrow/.default=black}

\draw[very thick,-] (0,0) -- (14,0);
\draw[very thick,-] (0,-0.02) -- (0,5.5);
\draw[thick,-] (0,4) node[left] {$k=4$} -- (14,4);

%\draw[help lines,-] (0,0) -- (13.5,0);
%\draw[help lines,-] (0,4) node[left] {$k=4$} -- (13.5,4);

\draw[big arrow] (0,0) -- (14,0) node[below right] {$i$};

\draw[big arrow] (0,-0.02) -- (0,5.5) node[left] {$j$};

% fill and dashed line
% fill=white, yellow, red, etc.

%\draw[ultra thick,dashed,fill=yellow,fill opacity=0.6](0,1)--(1,0)--(2,1)--(1,2)--(0,1);
%\draw[ultra thick,dashed,fill=yellow,fill opacity=0.6](4,1)--(6,3)--(7,2)--(8,3)--(9,2)--(11,4)--(13,2)--%(13,0)--(12,1)--(11,0)--(10,1)--(9,0)--(8,1)--(7,0)--(6,1)--(5,0)--(4,1);

\draw[thick,dashed,fill=white,fill opacity=0](0,1)--(1,0)--(2,1)--(1,2)--(0,1);
\draw[thick,dashed,fill=white,fill opacity=0](4,1)--(6,3)--(7,2)--(8,3)--(9,2)--(11,4)--(13,2)--(13,0)--(12,1)--(11,0)--(10,1)--(9,0)--(8,1)--(7,0)--(6,1)--(5,0)--(4,1);
\draw[thick,dashed](5,2)--(6,1)--(7,2)--(8,1)--(9,2)--(10,1)--(11,2)--(12,1)--(13,2);
\draw[thick,dashed](10,3)--(11,2)--(12,3);

% path from m=1 to n=2
\draw[ultra thick,purple,-](0,1)node[left] {{\color{black} $m=1$}}--(1,2)--(3,0)--(6,3)--(7,2)--(8,3)--(9,2)--(11,4)--(13,2)node[above right] {\color{black} $n=2$};

% initial point & terminal point
\fill (0,1) circle (2.2pt);
\fill (13,2) circle (2.2pt);

\end{tikzpicture}
}
\caption{\small{A typical path [solid (purple) meander] for $k=4$, starting at $m=1$ and ending at $n=2$, of length $l=13$ steps. The area under it is 27.5 plaquettes and subtracting the area of the triangular `sierra' at the bottom $l/2=6.5$ we obtain $A=21$. Relevant to the conventions after section 4.1, its length can also be expressed as $l=6.5$ double-steps and $A=21/2=10.5$ `diamonds'
formed by the dashed lines (there is a half-diamond at the last step of the path).}}\label{Dyck path}
\end{figure}
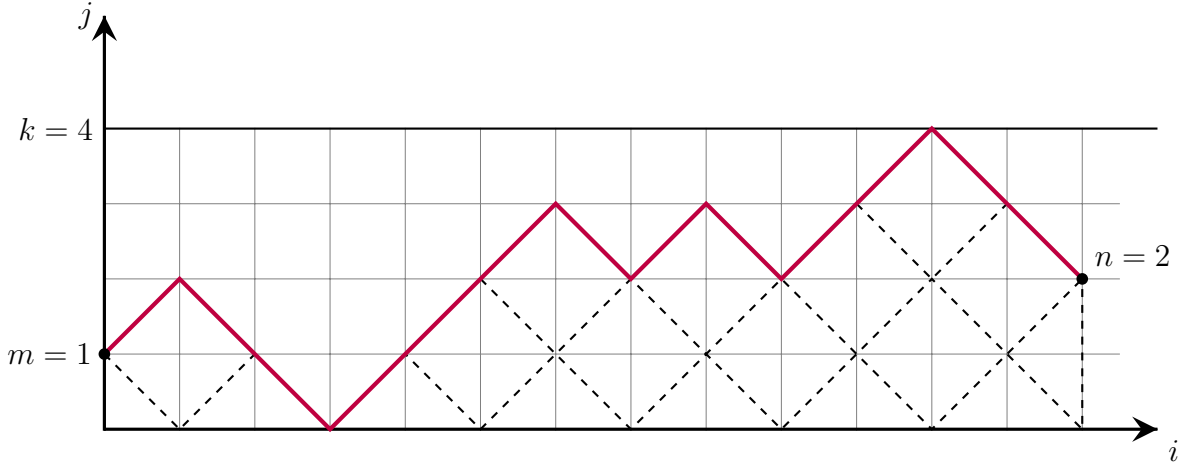}

We will view the Hamiltonian $H_k$ as generating random walks on a square lattice on the first quadrangle of the plane
consisting of points $(i,j)$, $i = 0,1,2,\dots$, $j = 0,1,2,\dots k$, while also keeping track of the area between the walk
and the horizontal axis (see fig.\ref{Dyck path}).
The Hamiltonian represents the transition matrix of the corresponding walk.
Specifically, the action of $H_k$ on $\ket{j}$ produces the superposition $\q^j \ket{j+1} + \q^{j-1} \ket{j-1}$.
Mapping the vertical position $j$ to the Hilbert space element $\ket{j}$, we can interpret the action of $H_k$
as producing a unit vertical step either up to $\ket{j+1}$, or down to $\ket{j-1}$. We consider a single application of $H_k$
to correspond to a unit horizontal step $i \to i+1$. The vertical area under such a step $(i,j) \to (i+1,j\pm 1)$
measured in units of lattice plaquettes is $a= \bigl( j+ (j\pm 1)\bigr)/2 = j\pm\half$ for
an up or down step, respectively, and therefore the weighting factors $\q^j$ and $\q^{j-1}$ are $\q^{a-1/2}$.
The repeated application of $H_k$, then, produces a collection of paths moving either
up or down by one unit in each of the $l$ horizontal steps, each path weighted by a factor arising from the products of
the above coefficients in each step; that is, by a factor $\q^{a - l/2}$ where $a$ is the total area under the path.
We will call $A = a- l/2$ the ``area'' of the walk (in a slight abuse of the term). It can be visualized as the area vertically under the walk reduced by the
area of the ``sierra'' at the bottom (see figure 1).
%Specifically, mapping the vertical position to the Hilbert space element $\ket{j}$,
%we will consider that $H_k$ acting on it generates one horizontal step $i \to i+1$. 
%Under a single action of $H_k$ on $\ket{j}$ the state
%will go either one step up to $\ket{j+1}$, and will also be multiplied by $\q^j$, or one step down to $\ket{j-1}$,
%multiplied by $\q^{j-1}$. This creates two possible walks, and the state branching process repeats in each step
%(successive application of $H_k$) leading to a collection of walks.
%The state $\ket{0}$ (corresponding to lattice
%points $(s,0)$) constitutes a `floor' and the state $\ket{k}$  (corresponding to points $(s,k)$) to a `ceiling' for the walk.

States $\ket{0}$ and $\ket{k}$ (corresponding to lattice points $(i,0)$ and $(i,k)$) constitute a `floor' and  a `ceiling'.
Walks starting and ending at the floor state $\ket{0}$ are called height-restricted Dyck paths and have been studied
extensively in the mathematics literature (the restriction refers to the existence of the ceiling $\ket{k}$). 
An object of special interest is the generating function of walks of $l$ steps and area $A$, starting at fixed height
$m$ and ending at fixed height $n$; that is,
\be
G_{k,mn} (\z,\q) = \sum_{A,l=0}^\infty \z^l \, \q^A N_{k,mn;l,A}
\ee
with $N_{k,mn;l,A}$ the number of walks with the given parameters. The correspondence of such walks with the action
of $H_k^l$, described above,
makes it clear that the $m,n$ matrix element of $H_k^l$ reproduces the sum of walks weighted by their area:
\be
\bra{m} H_k^l \ket{n} = \sum_{A=0}^\infty \q^A N_{k,mn;l,A}
\ee
and the full generating function becomes a matrix element of the propagator
\be
G_{k,mn} (\z,\q) = \sum_{l=0}^\infty \z^l \bra{m} H_k^l \ket{n} = \bra{m} (1- \z H_k )^{-1} \ket{n}
\ee
(we assumed small enough $|\z|$ and $|\q|$ for convergence of the sum). It is clear from this form that the generating
function satisfies the convolution property
\be
G_{k,mn} (\z,\q) = \sum_{l=0}^\infty G_{k,ml} (\z,\q) G_{k,ln} (\z,\q)
\ee
Further, by the symmetry of $H_k$, $G_{k,mn} (\z,\q) = G_{k,nm} (\z,\q)$.

In the sequel we will demonstrate that the above generating function can be expressed in terms of a ratio of determinants,
evaluate these determinants, and make
a connection between such walks and generalized quantum exclusion statistics of order 2. This connection leads to the
derivation of an alternative, explicit form of the generating function, and opens the path to further generalizations.

\section{Determinant formulae}\label{detsec}

Our goal is the evaluation of the matrix elements of the `propagator' matrix $(1-\z H_k )^{-1}$ that appear
in the generating function.

\subsection{Basic result}

Define the secular matrix
\be
D_k (\z,\q) = 1- \z H_k
=
\begin{pmatrix}
1& -\z  & 0 & 0\;\; \cdots & \bb\bb 0 & \bb 0 \\
-\z & 1 &\bb -\z \q & 0\;\; \cdots & \bb\bb 0 & \bb 0 \\
0 & -\z \q & 1 &-\z \q^2 \cdots & \bb\bb 0 & \bb 0 \\
\vdots & \vdots & \vdots & \;\;\; \ddots & \vdots &\bb \bb \vdots \\
0 & 0 & 0 & 0 \; \cdots & \bb 1 &\bb -\z \q^{k-1} \\
0& 0 & 0  &0 \;\cdots &\bb\bb -\z \q^{k-1}& 1 \\
\end{pmatrix}
\label{Dk}\ee
and its determinant and matrix elements of its inverse (generating function)
\be
F_k (\z,\q) = \det D_k (\z,\q) ~,~~~
G_{k,mn} (\z,\q) = \bra{m} D_k (\z,\q)^{-1} \ket{n}
\ee
Clearly $F_0 (\z,\q)=1$. For later convenience we also define 
$F_{-1} (\z,\q) =1$, and will adopt the convention that indices $\infty$ and $00$ are omitted;
%, while $kk$ becomes overbar;
that is,
\be
F_\infty = F ~,~~ G_{\infty,mn}  = G_{mn} ~ ,~~ G_{k,00}  = G_k ~,~~
%G_{k,kk}  = {\overline G}_k ~ ,~~ 
G_{\infty,00}  = G 
\ee

To calculate $G_{k,mn}$ we use the standard formula giving the elements of the inverse of a matrix in terms of its
complements. Applied to matrix $D_k (\z,\q)$ it yields
\be
\bra{m} D_k (\z,\q)^{-1} \ket{n}  = (-1)^{m-n} \; {\det D_k (\z,\q)_{(nm)} \over \det D_k (\z,\q)}
\ee
where the complement $D_k (\z,\q)_{(nm)}$ is the matrix $D_k (\z,\q)$ with the $n^{\text {th}}$ row and $m^{\text {th}}$ column removed.
The denominator in the right-hand side is $F_k (\z,\q)$. So
it remains to calculate the determinant of $ D_k (\z,\q)_{(nm)}$.

This determinant can be related to
simple secular determinants. First, observe that the secular matrix with its first $n$ rows
and columns truncated, denoted $D_k (\z,\q)_{[n]}$, is related to the secular matrix for a reduced $k$. Specifically,
\be
D_k (\z,\q)_{[n]} \bb = \bb \begin{pmatrix}
1& -\z \q^{n} & 0\;\; ~~~\cdots & \bb\bb\bb 0 & \bb\bb 0 \\
-\z \q^{n} & 1 &\bb -\z \q^{n+1}  \cdots & \bb\bb\bb 0 & \bb\bb 0 \\
0 & -\z \q^{{n+1}} & 1 ~~~ \cdots & \bb\bb\bb 0 & \bb\bb 0 \\
\vdots & \vdots  & \;~~ \; \ddots & \bb\bb\vdots &\bb \bb\bb \vdots \\
0 & 0 &  0 \; ~~~\cdots & \bb\bb 1 &\bb\bb -\z \q^{k-1} \\
0& 0  &0 \;~~~\cdots &\bb\bb -\z \q^{k-1}&\bb 1 \\
\end{pmatrix}
= D_{k-n} (\z \q^n , \q)
\label{trunc}
\ee
Assuming $m\le n$ (which is adequate, by the symmetry of $G_{k,mn} (\z,\q)$), the complement $D_k (\z,\q)_{(nm)}$ becomes
block-triangular (we leave it to the reader to visualize this) of the form
\be
D_k (\z,\q)_{(nm)} = 
\begin{pmatrix}
D_{m-1} (\z,\q) & 0 & 0 & \\
  &  &  \\
A & Q &0 \\
  &  &  \\
B & C &  \hskip 0.2cm D_k (\z,\q)_{[n+1]}\bb\bb \\
\end{pmatrix}
\ee
and therefore
\be
\det D_k (\z,\q)_{(nm)} = \det \bb D_{m-1} (\z,\q)\, \det\bb Q \, \det\bb D_k (\z,\q)_{[n+1]}
\ee
The form of $A,B,C$ is unimportant ($B$ mostly vanishes). $Q$, on the other hand, is a lower-triangular matrix
with diagonal elements $-\z \q^{m} , -\z \q^{m+1} , \dots , -\z \q^{n - 1}$.
Therefore,
\be
\det Q = (-\z)^{n-m} \, \q^{(n - m)(n+m-1) \over 2}
\ee
Putting everything together, and using (\ref{trunc}), we finally obtain
\be
G_{k,mn} (\z,\q) = \z^{n-m}\,  \q^{(n - m)(n+m-1) \over 2} \, {F_{m-1} (\z,\q) \, F_{k-n-1} (\z \q^{n+1} ,\q) \over F_k (\z,\q)}
~,~~ n\ge m
\label{typara}\ee
This is our main result. Its basic building block is the determinant of the secular matrix. We point out that expressions
of the length generating function ($\q=1$) in terms of ratios of determinants of the transition matrix have appeared before in the literature \cite{BM,Bach},
but our expression for the length and area generating function for arbitrary meanders as a rational expression of
subdeterminants is, to the best of our knowledge, new.

We also note that ``weighted'' generating functions assigning different weights to different kinds of steps have been considered.
In our case, this amounts to a trivial modification: a walk with $N_u$ up-steps and $N_d$ down-steps satisfies
\be
N_u + N_d = l ~,~~ N_u - N_d = n-m
\ee
Therefore, the generating function weighing up-steps and down-steps with factors $\z_u$ and $\z_d$, respectively, is
simply
\be
G_{k,mn} (\z_u,\z_d ;\q) = (\z_u \z_d^{-1})^{n-m \over 2}\, G_{k,mn} \big((\z_u \z_d )^{1/2},\q\big)
\ee
%\be
%G_{k,mn} (\z_u,\z_d ;\q) = \left(\z_u \over \z_d \right)^{n-m \over 2} G_{k,mn} \big((\z_u \z_d )^{1/2},\q\big)
%\ee
In the next section we will connect the secular determinant to quantum exclusion statistics, which will also
provide a method for calculating it and finding a more explicit form.

\subsection{Special cases}

It is useful to write this result for a few interesting special cases. For ``diagonal'' height-restricted paths, starting and ending at
the same height $m=n$, we have
\be
G_{k,nn} (\z,\q) = {F_{n-1} (\z,\q) F_{k-n-1} (\z \q^{n+1} ,\q) \over F_k (\z,\q)}
\ee
In particular, for excursions starting and ending at $n=0$, $G_{k,00} (\z,\q) = G_k (\z,\q)$ is
\be
G_{k} (\z,\q) = {F_{k-1} (\z \q ,\q) \over F_k (\z,\q)}
\ee
Expressions similar to the above for the length generating function of excursions ($\q \bb=\bb 1$) have been derived in \cite{BM} and for the
length and area generating function in \cite{BIP}. For the `dual' paths starting and ending at the ceiling we obtain
\be
{G}_{k,kk} (\z,\q) = {F_{k-1} (\z ,\q) \over F_k (\z,\q)}
\ee
Finally, for unrestricted Dyck excursions, starting and ending at $n=0$, we have
\be
G (\z,\q) = {F (\z \q ,\q) \over F (\z,\q)}
\ee

\subsection{Duality and recursion relations}

Walks are obviously symmetric under vertical reflection around the median line at $k/2$, leading to the mapping
$\ket{n} \to \ket{k-n}$. A consequence of this symmetry is that the secular matrix and determinant satisfy
the duality relation
\be
D_k (\z \q^{k-1} ,\q^{-1}) = \sigma D_k (\z,\q)\, \sigma ~,~~~ F_k (\z \q^{k-1} , \q^{-1} ) = F_k (\z,\q)
\ee
where $\sigma_{mn} = \delta_{m+n,k}$ is the reflection matrix.
This leads to the corresponding duality relation for generating functions
\be
G_{k,mn} (\z,\q) = G_{k;k-m,k-n} (\z \q^{k-1}, \q^{-1} )
\ee

Several generating function recursion relations can be deduced directly from the form itself of (\ref{typara}). For instance,
%\be
%G_{k,mn} (\z,\q)  = \z \q^{n-1}\, G_{k;m,n-1} (\z,\q) \, G_{k-n} (\z \q^n ,\q)
%= \z \q^{m-1}\, G_{k;m+1,n} (\z,\q) \, {\overline G}_{k-m} (\z,\q)
%\ee
\bea
G_{k,mn} (\z,\q)  &=& \z \q^{n-1}\, G_{k;m,n-1} (\z,\q) \, G_{k-n} (\z \q^n ,\q) ~~~~~~~~~~~~(m<n) \nonumber \\
&=&\, \z \q^{l}\; G_{k;l+1,n} (\z,\q) \, {G}_{l;m,l} (\z,\q)~\,~~~~~~~~~~~~~~~~~(m \le l < n) \label{firstpass}
%&=& \z \q^{n-1}\, G_{k;m,n-1} (\z,\q) + \z \q^n\,  G_{k;m,n+1} (\z,\q) ~~~(m<n<k) \nonumber
%&=& \z^{-1} \q^_-n} G_{k;m,n+1} (\z,\q) \, G_{k-n-1} (\z \q^{n+1} ) ~~~~~~(n<k) \\
\eea
Further recursion relations can be derived by expanding $\det D_k (\z,\q)$ in terms of its top row.
We obtain
\be
F_k (\z,\q) = F_{k-1} (\z \q ,\q ) - \z^2 F_{k-2} (\z \q^2 , \q)
\label{ex}\ee
which leads to corresponding relations for $G_{k,nm} (\z,\q)$. Several such relations can be written, but we will
only present two: for generic paths, applying (\ref{ex}) to the term $F_{k-n-1} (\z \q^{n+1} ,\q)$ in (\ref{typara}) yields
%\be
%G_{k,nm} (\z,\q) = G_{k-1;n-1,m-1} (\z \q,\q) - \z^2 \q \, G_{k-2;n-2,m-2} (\z \q^2,\q) G_{k-1} (\z \q,\q) G_k (\z,\q)
%\ee  
\be
G_{k,mn} (\z,\q) = \z \q^{n-1}\, G_{k;m,n-1} (\z,\q) + \z \q^n\,  G_{k;m,n+1} (\z,\q) ~~~~~(m<n<k)
\ee
and for paths starting and ending at $n=0$, dividing (\ref{ex}) by $F_k (\z,\q)$ yields
\be
G_{k} (\z , \q) = 1 + \z^2 \, G_{k-1} (\z \q,\q) \, G_k (\z,\q) 
\label{concur}\ee
All the above recursion relations have clear geometric interpretations in terms of constructing (or deconstructing)
paths in terms of their parts (see, e.g., figure \ref{Geometric}).

{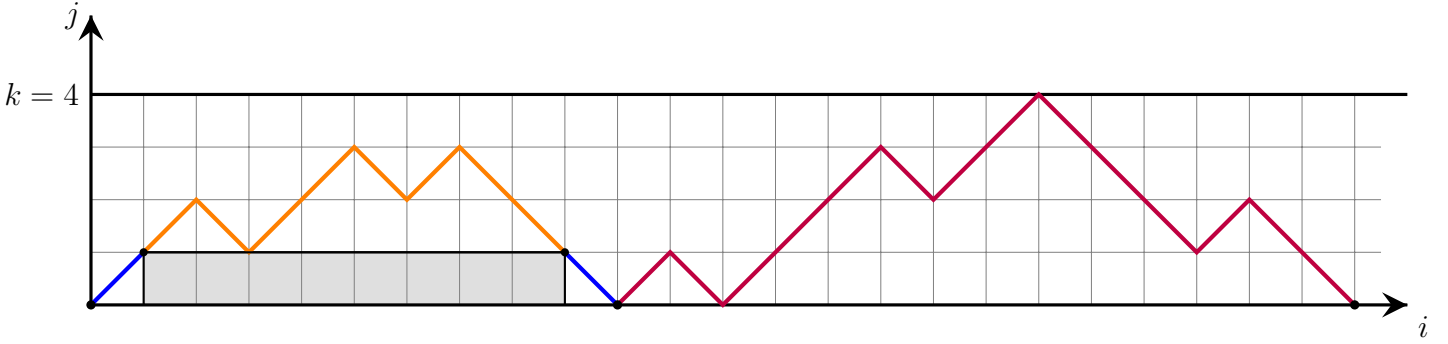
\begin{figure} \vskip -1.8cm
%\centering 
{\hskip -1.5cm
\begin{tikzpicture}[scale=0.7]
% square grid
\draw[help lines, gray] (0,-0.02) grid (24.5,4.01);

\tikzset{big arrow/.style={decoration={markings,mark=at position 1 with {\arrow[scale=3,#1,>=stealth]{>}}},postaction={decorate},},big arrow/.default=black}
% Different styles for arrows: >=to (default), stealth, latex
% Alternative type: \tikzset{big arrow/.style={decoration={markings,mark=at position 1 with {\arrow[scale=2,#1,>=to]{>}}},postaction={decorate},},big arrow/.default=black}

\draw[very thick,-] (0,0) -- (25,0);
\draw[very thick,-] (0,-0.02) -- (0,5.5);
\draw[very thick,-] (0,4) node[left] {$k=4$} -- (25,4);
\draw[thick,-] (1,1) -- (9,1);

\draw[big arrow] (0,0) -- (25,0) node[below right] {$i$};
\draw[big arrow] (0,-0.02) -- (0,5.5) node[left] {$j$};

% path from m=1 to n=2
\draw[ultra thick,orange,-](1,1)--(2,2)--(3,1)--(4,2)--(5,3)--(6,2)--(7,3)--(8,2)--(9,1);
\draw[ultra thick,blue,-](0,0)--(1,1); \draw[ultra thick,blue,-](9,1)--(10,0);
\draw[ultra thick,purple,-](10,0)--(11,1)--(12,0)--(15,3)--(16,2)--(18,4)--(21,1)--(22,2)--(24,0);
\draw[thick,-=0.5,fill=gray,fill opacity=0.25](1,0)--(1,1)--(9,1)--(9,0)--(1,0);

% initial point & terminal point
\fill (0,0) circle (2.5pt);\fill (1,1) circle (2.2pt);\fill (9,1) circle (2.2pt);
\fill (10,0) circle (2.5pt);\fill (24,0) circle (2.5pt);

\end{tikzpicture}
}\vskip -0.3cm
\caption{\small{A geometric interpretation of recursion relation (\ref{concur}) as a ``first passage'' equation. Any excursion 
(one of length 24 pictured above) can
be decomposed into a first-passage path returning to height $j=0$ for the {\it first time} [first (blue \& orange) part of path] and the remaining (purple) arbitrary
excursion. For paths of length at least two, the first-passage path has a first and a last step
[lower (blue) links]. The remaining upper (orange) part never dips below $j=1$ and can be interpreted as an excursion, but with a
ceiling reduced by 1 and an area increased by its length (shaded plaquettes).
% between $j=1$ and $j=2$).
Summing over all paths gives the full generating function $G_k (\z,\q)$ as the product of itself [second (purple) excursion] times
$G_{k-1} (\z \q,\q)$ [first (orange) excursion], the shift $\z \to \z\q$ accounting for the increase in the area, times $\z^2$
contributed by
the two lower (blue) links. The trivial path of zero length and area cannot be decomposed
and contributes the term 1 in (\ref{concur}). Relations (\ref{firstpass}) also admit a similar first-passage interpretation.}}\label{Geometric}
\end{figure}}

\section{Exclusion statistics connection}

In \cite{emeis} a connection was pointed out between two-diagonal matrices and quantum exclusion statistics
(for a review of exclusions statistics see \cite{gofel}). Specifically, the secular determinant of a matrix
with two nonzero off-diagonals, one (with elements $f_n$) just above the (vanishing) diagonal and the other
(with elements $g_n$) $g-1$ steps below the diagonal, is given by the {\it grand partition function} of noninteracting particles
of exclusion statistics $g$ with single-particle Boltzmann factors $s(n) \bb =\bb e^{-\beta \varepsilon_n}$, $n=0,1,2,\dots$
($1/\beta = k_{_B} T$ as usual)
and fugacity parameter $x = e^\mu$ given by
\be
s (n) = g_n \, f_n\,  f_{n+1} \cdots f_{n+g-2} ~,~~~ x = -\z^g
\label{ex2}\ee
$s(n)$, called the spectral function in \cite{emeis}, encodes the single-particle energy spectrum.

The matrix $H_k$ in (\ref{hamk}) is of the above form, with $g=2$ and $f_n = g_n = \q^{n}$, for a spectral function
and fugacity
\be
%e^{-\beta \varepsilon_n} \equiv 
s (n) = \q^{2n} = e^{-2n \ln \q^{-1}} ,~~n=0,1,\dots,k\bb -\bb 1 ~;~~~ x = -\z^2
\label{spec}\ee
These are the Boltzmann factors for energy levels $\varepsilon_n =  n \beta^{-1} \ln(\q^{-2})$, that is, a finite equidistant spectrum, which we will call a truncated 
harmonic oscillator spectrum (levels range from $\varepsilon_0$ to $\varepsilon_{k-1}$). 
The determinant $F_k (\z,\q)$ is the grand partition function of noninteracting  exclusion-2 particles in this spectrum, obeying
the exclusion principle that no two particles can be in the same {\it or neighboring} levels.

\subsection{A change in conventions}

This is an opportune moment to make a change of conventions for the variables $\q$ and $\z$ dual to the area and length
of the walks. The spectrum and fugacity parameter in (\ref{spec}) suggest the use of
parameters
\be
z = \z^2 ~,~~ q = \q^2 ~~~~{\text {and~thus}}~~~~ s(n) = q^n ~,~~ x = -z
\label{specc}\ee
as more natural. Adopting them in the previous sections would have made intermediate expressions rather awkward,
involving square roots, but they become
a better choice in the sequel. This change amounts to measuring the area in terms of diamonds, of  area double that
of an elementary plaquette, and the length in terms of double steps. Paths starting and ending at different heights
$m\neq n$ can have {\it half}-integer area and length in the new conventions, but this is not a serious drawback.

Renaming $F_k (z^2, q^2)$ and $G_{k,mn} (z^2 , q^2)$ simply $F_k (z, q)$ and $G_{k,mn} (z , q)$, all the basic
formulae of the previous sections remain essentially unchanged, modified only when explicit factors of $\z$ or
$\q$ are involved. We list here three main formulae:
\bea 
G_{k,mn} (z,q) &=& z^{{n-m}\over 2}\,  q^{(n - m)(n+m-1) \over 4} \, {F_{m-1} (z,q) \, F_{k-n-1} (z q^{n+1} , q) \over F_k (z,q)} ~,~~ n\ge m\nonumber \\
F_k (z,q) &=& F_{k-1} (z q ,q ) - z F_{k-2} (z q^2 , q) \label{newG} \\
G_{k} (z , q) &=& 1 + z \, G_{k-1} (z q,q) \, G_k (z,q)  \nonumber
\eea

\subsection{Grand partition functions}

The determinant $F_k (z,q)$ is expressed as the grand partition function of noninteracting  exclusion-2 particles in the
single-particle spectrum $\varepsilon_n = n \beta^{-1} \ln (q^{-1}) = \varepsilon n$:
\be
F_k (z,g) = \sum_{N=0}^{\lfloor{k+1 \over 2}\rfloor} (-z)^N Z_{k,N}^{(2)} (q)
\label{FkZ2}\ee
with $N$ the number of particles and $Z_{k,N}^{(2)}$ the partition function of $N$ exclusion-2 particles in the above
spectrum. The $N$-particle ground state consists of levels $n\bb =\bb 0,2,4,\dots,2N-2$ occupied by a single particle,
with Boltzmann factor $q^{N(N-1)}$. Excited states correspond to $N$ occupied levels with at least
one empty level between them.

For an equidistant energy spectrum a standard bosonization argument allows to write $N$-body
partition functions in terms of bosonic partition functions. The standard case of bosonization is for relativistic massless fermions on the periodic line, for which the energy spectrum is also equidistant, but it applies to our exclusion statistics
particles as well. The basic bosonization trick is to reduce the gap between successive occupied levels by $g$, which then ``collapses'' the state into a bosonic state but also adds a constant to the energy. Specifically, the energy of a state
of $N$ particles of exclusion $g$ can be expressed in terms of the successive occupied level $\ell_i$ as
\be
E = \sum_{i=1}^N \varepsilon \ell_i ~,~~~ 0\le \ell_i + 2 \le \ell_{i+1} \le k-1
\ee
The condition $\ell_i + 2 \le \ell_{i+1}$ enforces $g=2$ exclusion, guaranteeing that neighboring levels cannot
both be occupied. Redefining $\ell_i = {l}_i - 2(i-1)$ the energy becomes
\be
E = \varepsilon {N(N-1)} + \sum_{i=1}^N \varepsilon {l}_i ~,~~~0\le  {l}_i \le{l}_{i+1} \le k-1-2(N-1)
\ee
That is, the energy of noninteracting bosons at levels $l_i$ with a zero-point energy $\varepsilon N(N-1)$
and a reduced number of levels $k-2(N-1)$. This allows us to rewrite
the grand partition function (\ref{FkZ2}) in terms of {\it bosonic} partition functions
\be
F_k (z,q) = \sum_{N=0}^{\lfloor{k+1 \over 2}\rfloor} (-z)^N  q^{N(N-1)} Z_{k-2(N-1),N}^{B} (q)
\label{boso}\ee
For finite $k$ the above is a finite sum, truncating at $N=\lfloor{k+1 \over 2}\rfloor$.
The bosonic partition function of $N$ particles in $k$ oscillator states can be expressed as
%\bea
%Z_{k,N}^{B} (q) &=& \sum_{\{n_i =0\}}^\infty q^{\sum_{i=0}^{k-1} i \, n_i}\; \delta \bb\left( \sum_{i=0}^{k-1} n_i - N \right)
%\nonumber \\
% &=& \sum_{\{m_j =0\}}^\infty q^{\sum_{j=0}^N j m_j} \; \delta\bb \left( \sum_{j=0}^N m_j - k+1 \right)
%\label{ZB} \eea
%The top expression is an expansion in terms of the {\it occupation} numbers $n_i$ of single-particle states $i=0,1,\dots,k-1$,
%while the bottom one is an expansion in terms of the {\it excitation} numbers $m_j$ of the top $j$ particles ($m_0$ being
%simply the number of empty levels above the top particle).
\begin{subequations}
\bea
Z_{k,N}^{B} (q) 
&=& \sum_{\{n_i =0\}}^\infty q^{\sum_{i=0}^{k-1} i \, n_i}\; \delta \bb\left( \sum_{i=0}^{k-1} n_i - N \right)
\label{ZBa}%\nonumber 
\\
&=& \sum_{\{m_j =0\}}^\infty q^{\sum_{j=0}^N j m_j} \; \delta\bb \left( \sum_{j=0}^N m_j - k+1 \right) 
\label{ZBb}
\\
&=& {[k+N-1]!_q \over [N]!_q  \, [k-1]!_q} = {k+N-1 \choose N}_{\bb\bb q} 
\label{ZBc}
 \\
&=& \prod_{j=1}^N {1-q^{j+k-1} \over 1-q^j}
=  \prod_{j=1}^{k-1} {1-q^{j+N} \over 1-q^j}\label{ZBd}
%= {(q;q)_{N+k-1} \over (q;q)_N \, (q;q)_{k-1}}
\eea
\end{subequations}
We deliberately gave several alternatives:
(\ref{ZBa}) is an expansion in terms of the {\it occupation} numbers $n_i$ of single-particle states $i=0,1,\dots,k-1$,
while (\ref{ZBb})  is an expansion in terms of the {\it excitation} numbers $m_j$ of the top $j$ particles ($m_0$ being
simply the number of empty levels above the top particle). These are related to the bosonic levels $l_i$ as
\begin{subequations}
\bea
m_i &=& l_{N-i+1} - l_{N-i} ~,~~~ i=1,2,\dots,N \label{MNLa}\\
n_i &=& \sum_{j=1}^N \delta(l_j -i) ~,~~~ i=0,1,\dots,k-1 \label{MNLb}\\
\Rightarrow ~~~n_i &=& \sum_{j=1}^N \delta \left( \sum_{s=0}^{j-1} m_{N-s} -i \right) \label{MNLc}
\eea
\end{subequations}
From the above we deduce
\be
E = \sum_{i=0}^{k-1} i n_i = \sum_{j=1}^N j m_j ~;~~~~
\sum_{i=0}^{k-1} n_i = N ~,~~~~ \sum_{j=1}^N m_j \le k-1
\ee
which, upon defining $m_0=-\sum_{j=1}^N m_j + k-1 \ge 0$ leads to (\ref{ZBa}) and (\ref{ZBb}).
Evaluation of the sums leads to the other two expressions for the grand partition function:
(\ref{ZBc}) expresses it as a $q$-deformation of the 
degeneracy of $N$ bosons in $k$ states, while (\ref{ZBd}) gives the most explicit formula.
The $q$-deformed factorial (and corresponding $q$-choose symbol) in (\ref{ZBc}) is defined as
\be
[n]!_q = \prod_{j=1}^n [j]_q = \prod_{j=1}^n {1-q^j \over 1-q}
\ee
It is clear that $Z_{k,N}^{B} (q)$ obeys the duality $N \leftrightarrow k-1$
\be
Z_{k,N}^{B} (q) = Z_{N+1,k-1}^{B} (q)
\ee
valid specifically for the equidistant spectrum. Using the second expression in (\ref{ZBd}), the
grand partition function $F_k (z,q)$ (\ref{boso}) becomes
\be
F_k (z,q) = \sum_{N\hskip-0.03cm =0}^{\lfloor{k+1 \over 2}\rfloor} (-z)^N q^{N(N-1)}  {k-N+1 \choose N}_{\bb\bb q}  =
\sum_{N\hskip-0.03cm =0}^{\lfloor{k+1 \over 2}\rfloor} \prod_{j=1}^N (-z)\, {q^{N-1} - q^{j+k-N} \over 1-q^j}
\label{Fk}\ee
%\be
%F_k (z,q) = \sum_{N=0}^\infty (-z)^N  q^{N(N-1)} 
%\sum_{\{m_j =0\}}^\infty q^{\sum_{j=0}^N j m_j} \; \delta\bb \left( \sum_{j=0}^N m_j +2N - k-1 \right)
%\label{Fk}\ee
which expresses $F_k (z,q)$ as a $q$-deformed Fibonacci polynomial in $-z$.
For $k=0$ only the $N=0$ term survives and we obtain $F_0 (z,q) =1$.

The $F_k (z,q)$ for various values of $k=0,1,2,\dots$ can be repackaged into a height-generating function 
$H(w,z,q) = \sum_k w^k F_k (z,q)$. Using (\ref{ZBb}) we obtain
%\be
%H(w,z,q) = \sum_{k=0}^\infty w^k F_k (z,q) = {1\over w} \left[ -1 + 
%\sum_{N=0}^\infty (-w^2 z)^N \,{q^{N(N-1)}\over (w ; q)_{N+1} } \right]
%\ee
\be
H(w,z,q) = \sum_{k=0}^\infty w^k F_k (z,q) = -{1\over w} +{1\over w} 
\sum_{N\hskip-0.03cm =0}^\infty (-w^2 z)^N \,{q^{N(N-1)}\over (w ; q)_{N+1} }
\ee
%\be
%H(w,z,q) = \sum_{k=0}^\infty w^k F_k (z,q) = {1\over 1-w} \left[ 1+ {1 \over w} 
%\sum_{N=1}^\infty (-w^2 z)^N \,{q^{N(N-1)}\over (w q ; q)_N } \right]
%\ee
with the $q$-Pochhammer symbol $(a;q)_n$ defined as
\be
(a;q)_n = \prod_{j=0}^{n-1} (1- a q^j )
\ee

Finally, in order to make the exclusion statistics connection more relevant and intuitive, we point out
that the recursion relation (\ref{newG}) admits a clear statistical interpretation: $F_k (z,q)$ in the left-hand side is the grand partition function for levels $1,2,\dots ,k$. Focusing our attention on the first level $k=1$, it could be either empty or filled.
If it is empty, the grand partition function is the one for levels $2,\dots,k$, which is $ F_{k-1} (zq ,q )$; if it is full, it 
contributes a Boltzmann factor $-z$ times the grand partition function for levels $3,\dots,k$, since level $2$
by the exclusion principle cannot be filled, which is $F_{k-2} (z q^2 , q)$.
Summing the two types of configurations gives the right-hand side.

\section{Application to walks}

In the previous section we obtained relatively explicit formulae for $F_k (z,q)$, and therefore for $G_{k,mn} (z,q)$.
Their form, however, in terms of sums of products, makes them rather unwieldy. In this section we will provide simpler
formulae for the
logarithm of the generating function $G_{k,mn}$ by making further use of the exclusion statistics connection.
To make the analysis more accessible, we will first treat the case of regular Dyck paths, focusing on the generating
function $G(z,q)$ of unrestricted excursions (floor-to-floor paths), and subsequently we will generalize to
height-restricted paths and arbitrary $G_{k,mn}$.

\subsection{Dyck excursions, $k=\infty$}

For the case of unrestricted Dyck paths, assuming $|q|<1$, (\ref{ZBd}) simplifies to
\be
Z_N^B (q) = {1 \over (1-q)^N [N]!_q} = {1 \over (q ; q )_N}
\ee
leading to the grand partition function
\be
F (z,q) =\sum_{N=0}^\infty (-z)^N Z_{N}^{(2)} (q) = \sum_{N=0}^\infty (-z)^N  \;{q^{N(N-1)} \over (q;q)_N}
\ee
Focusing on the floor-to-floor generating function $G(z,q)$, its expression becomes
\be
G (z,q) = {F (z q, q) \over F (z,q)}  = 
{\sum_{N=0}^\infty (-z)^N  \;{q^{N^2} \over (q;q)_N} \over \sum_{N=0}^\infty (-z)^N  
\;{q^{N(N-1)} \over (q;q)_N}}
\label{GG}\ee

The ratio of the two partition functions appearing in (\ref{GG}) can be further reduced
by using the cluster decomposition of the grand potential of the exclusion system.
Cluster coefficients $b_a$, $a=1,2,\dots$, are defined in terms of the expansion
of the grand potential $\ln Z(x)$ in terms of the fugacity $x$
\be
\ln Z(x) = \ln\left(\sum_{N=0}^{\infty}x^N Z_N  \right)=\sum_{a=1}^{\infty} x^a\, b_a
\ee
or, in our case,
\be
\ln F(z,q) = \ln\left(\sum_{N=0}^{\infty} (-z)^N Z_N^{(2)} (q)  \right)=\sum_{a=1}^{\infty} (-z)^a\, b_a
\label{Fbn}\ee
In \cite{nous,emeis} the general expression of these cluster coefficients was derived. For exclusion-2 statistics with
spectral parameter $s(n)$ it reads
\be 
b_a = (-1)^{a-1}\hskip -0.3cm \sum_{l_1, l_2, \ldots, l_{j}\atop { \rm compositions}\;{\rm of}\;a} \hskip -0.4cm 
c_2 (l_1,l_2,\ldots,l_{j} )\sum _{m=0}^{\infty} \prod_{i=1}^{j} s^{l_{i}} (m+i-1) 
\label{bn}\ee
(Compositions are partitions where the order of terms also matters.) 
The combinatorial coefficients $c_2 (l_1,l_2,\ldots,l_{j} )$ for exclusion-2 statistics are
\be
{c_2 (l_1,l_2,\ldots,l_{j})} 
= {1 \over l_1} \prod_{i=1}^{j-1} {l_i + l_{i+1}-1  \choose l_{i+1}}
%=\prod_{i=1}^{j-1} {l_i + l_{i+1}  \choose l_i}{1 \over l_j} 
= {{\prod_{i=1}^{j-1} (l_i + l_{i+1} -1)! \over \prod_{i=2}^{j-1} (l_i -1 )! \prod_{i=1}^j { l_i!}} }
\label{cl}\ee

Application of formulae (\ref{Fbn},\ref{bn}) for the specific spectral function (\ref{specc}) yields
\bea
\ln F(z,q) &=& \sum_{a=1}^{\infty} -z^{a} \hskip -0.4cm \sum_{l_1, l_2, \ldots, l_{j}\atop { \rm compositions}\;{\rm of}\;a} \hskip -0.4cm c_2 (l_1,l_2,\ldots,l_{j} )
\sum _{m=0}^{\infty} q^{\sum_{i=1}^{j} (m+i-1) l_{i}} \nonumber \\
\hskip -0.2cm\Bigl(\text{using}~\sum_{i=1}^j l_i = a \Bigr)~~~~&=& \sum_{a=1}^{\infty} -z^{a} \hskip -0.3cm \sum_{l_1, l_2, \ldots, l_{j}\atop { \rm compositions}\;{\rm of}\;a} \hskip -0.4cm c_2 (l_1,l_2,\ldots,l_{j} )\sum _{m=0}^{\infty} 
q^{ m a} \; q^{\sum_{i=1}^j (i-1) l_i}\nonumber \\
&=& \sum_{a=1}^{\infty} -{z^{a} \over 1-q^{a}} \hskip -0.3cm \sum_{l_1, l_2, \ldots, l_{j}\atop { \rm compositions}\;{\rm of}\;a} \hskip -0.4cm c_2 (l_1,l_2,\ldots,l_{j} ) ~q^{\sum_{i=1}^j (i-1) l_i}
\label{logF}\eea
and therefore
\bea
\ln G(z,q) &=& \ln F(z q ,q) - \ln F(z,q) \nonumber \\
&=& \sum_{a=1}^{\infty} z^{a} \hskip -0.3cm \sum_{l_1, l_2, \ldots, l_{j}\atop { \rm compositions}\;{\rm of}\;a} \hskip -0.4cm c_2 (l_1,l_2,\ldots,l_{j} ) ~q^{\sum_{i=1}^j (i-1) l_i} \label{lnG}\\
&=&  \sum_{a=1}^{\infty} z^{a} \, p_a (q) \nonumber
\eea
$p_a (q)$ is a polynomial in $q$ of degree $a(a-1)/2$ given by
%\be
%p_n (\q) = \hskip -0.5cm\sum_{l_1, l_2, \ldots, l_{j}\atop { \rm compositions}\;{\rm of}\;n} \hskip -0.4cm c(l_1,l_2,\ldots,l_{j} ) ~\q^{2\sum_{i=1}^j (i-1) l_i}
%\ee
\be
p_a (q) = \hskip -0.5cm\sum_{l_1, l_2, \ldots, l_{j}\atop { \rm compositions}\;{\rm of}\;a} \hskip -0.2cm
{{\prod_{i=1}^{j-1} (l_i + l_{i+1} -1)! \over \prod_{i=2}^{j-1} (l_i -1 )! \prod_{i=1}^j { l_i!}} }~q^{\sum_{i=1}^j (i-1) l_i}
\ee
The above manifests that the order $z^{a}$ term of $\ln G(z,q)$ is
a polynomial in $q$, in spite of the infinite sums in (\ref{GG}). Therefore, since $a(a-1)$ is a convex function of $a$,
the order $z^{a}$ term in $G(z,q)$ is also a polynomial $P_a (q)$ of the same degree $a(a-1)/2$.
This is consistent with its interpretation as the generating function of Dyck paths with $l=2a$ steps: the maximal area
is that of a triangular ``roof'' path with total area $a^2/2$, and subtracting half the length $a/2$ we end up with
$A=a(a-1)/2$.

\subsection{Rise-restricted meanders: finite $k$, general $m,n$}

For finite $k$ the explicit form of $F_k (z,q)$ is given by (\ref{Fk}), a finite but unwieldy sum. The calculation of
$G_{k,mn} (z,q)$, however, through cluster coefficients is tractable
and proceeds along similar lines as for $k=\infty$. (\ref{Fbn}) becomes
\be
\ln F_k (z,q) = \ln\left(\sum_{N=0}^{\infty} (-z)^N Z_{k,N}^{(2)} (q)  \right)=\sum_{a=1}^{\infty} (-z)^a b_{k,a}
\label{nbn}\ee
The cluster coefficients $b_{k,a}$ are now given by
\be 
b_{k,a} = (-1)^{a-1}\hskip -0.3cm \sum_{l_1, l_2, \ldots, l_{j} ;\, j\le k\atop { \rm compositions}\;{\rm of}\;a} \hskip -0.4cm 
c_2 (l_1,l_2,\ldots,l_{j} )\sum _{m=0}^{k-j} \prod_{i=1}^j q^{l_i (m+i-1)}
\ee
with the same coefficients $c_2 (l_1,\dots,l_j )$ as in (\ref{cl}).
Note that, now, only compositions with up to $k$ terms are included in the summation, and the $m$-summation is truncated.

A similar calculation as in (\ref{logF}) for $F_k (z,q)$ leads to
\be
\ln F_k (z,q) = \sum_{a=1}^{\infty} -{z^{a} \over 1-q^{a}} \hskip -0.1cm \sum_{l_1, l_2, \ldots, l_{j} ;\, j\le k\atop { \rm compositions}\;{\rm of}\;a} \hskip -0.4cm c_2 (l_1,l_2,\ldots,l_{j} ) \left[1- q^{(k-j+1)a} \right] q^{\sum_{i=1}^j (i-1) l_i}
\label{lnFk}\ee
The calculation of $\ln G_{k,mn} (q,z)$, using (\ref{typara}), proceeds along similar lines as in the $k=\infty$, $m=n=0$
case. Using the fact that the range of $j$ in the summation in (\ref{lnFk}) can be extended to $k+1$, since the corresponding
term vanishes, terms can be collected to express the final result as
%\bea
%\ln G_{k,mn} (z,q) &=& \ln F_{m-1} (z,q) +\ln F_{k-n-1} (z q^{n+1} ) - \ln F_k (\z,\q) \nonumber \\
%&=&  {{n-m}\over 2}\ln z +\,  {(n - m)(n+m-1) \over 4}\ln q\, + \sum_{a=1}^{\infty} z^{a}~ p_{k,mn;a} (q) \\
%\bb\text{with} ~~~~p_{k,mn;a} &=& \hskip -0.4cm\sum_{l_1, l_2, \ldots, l_{j} ;\, j\le k\atop { \rm compositions}\;{\rm of}\;n}
%\hskip -0.4cm c_2 (l_1,l_2,\ldots,l_{j} )\,{q^{\max\bb{(m-j,0)}a} - q^{\min\bb{(k-j+1,n+1)} a}  \over 1-q^a}
%\, q^{\sum_{i=1}^j (i-1) l_i} \nonumber 
%\eea
\bea
\hskip -0.3cm \ln G_{k,mn} (z,q) &=& \ln\bb\left(\bb z^{{n-m}\over 2} q^{(n - m)(n+m-1) \over 4}\bb\right)+\ln F_{m-1} (z,q) +\ln F_{k-n-1} (z q^{n+1},q ) - \ln F_k (z,q) \nonumber \\
&=&  {{n-m}\over 2}\ln z +\,  {(n - m)(n+m-1) \over 4}\ln q\, + \sum_{a=1}^{\infty} z^{a}~ p_{k,mn;a} (q)
\label{lnGk}\eea
%\be
%p_{k,mn;a} = \hskip -0.4cm\sum_{l_1, l_2, \ldots, l_{j} ;\, j\le k\atop { \rm compositions}\;{\rm of}\;a}
%\hskip -0.4cm c_2 (l_1,l_2,\ldots,l_{j} )\,{q^{\max\bb{(m-j,0)}a} - q^{\min\bb{(k-j+1,n+1)} a}  \over 1-q^a}
%\, q^{\sum_{i=1}^j (i-1) l_i}
%\ee
\be
\hskip -0.3cm\text{with} ~~~~~ p_{k,mn;a} = \hskip -0.4cm\sum_{l_1, l_2, \ldots, l_{j} ;\, j\le k\atop { \rm compositions}\;{\rm of}\;a}
\hskip -0.4cm c_2 (l_1,l_2,\ldots,l_{j} ) \, q^{\sum_{i=1}^j (i-1) l_i}\sum_{r={\max\bb{(m- j,0)}}}^{\min\bb{(k-j,n)} } q^{r a}
\ee
%In spite of the presence of the denominator, 
$p_{k,mn;a} (q)$ is a polynomial of degree
\be
{\text {degree}}~p_{k,mn;a} (q) = \left\{ \begin{matrix}
{a(a-1) \over 2}+ a n~~ & ,~~ a \le k-n\\
 & \\
{(k-n-1)(2a-k+n) \over 2}+a n~& ,~~a > k-n\\
\end{matrix} \right.
\label{degreek}\ee
$G_{k,mn} (z,q)$ is the exponential of (\ref{lnGk}). Therefore, the term of order $z^{a+(n-m)/2}$ is a polynomial in $q$
of the  same degree as $p_{k,mn;a} (q)$ times a prefactor $q^{(n-m)(n+m-1)/4}$.
This is consistent with its interpretation as the generating function of height-restricted Dyck paths: for paths with $2a+n-m$ steps and $a\le k-n$, the area is that of unrestricted paths, as paths of such length cannot exceed the ceiling.
For $n>k$, however, the path of maximal
area is a ``skew trapezoid tiled roof'' ascending from $m$, reaching the level $k$, and zigzagging $a-k+n$ times between
$k$ and $k-1$ before descending to $n$ (see fig.\ref{Tiled roof}).
It can be checked that such a path has an area consistent with the second entry in (\ref{degreek}).
{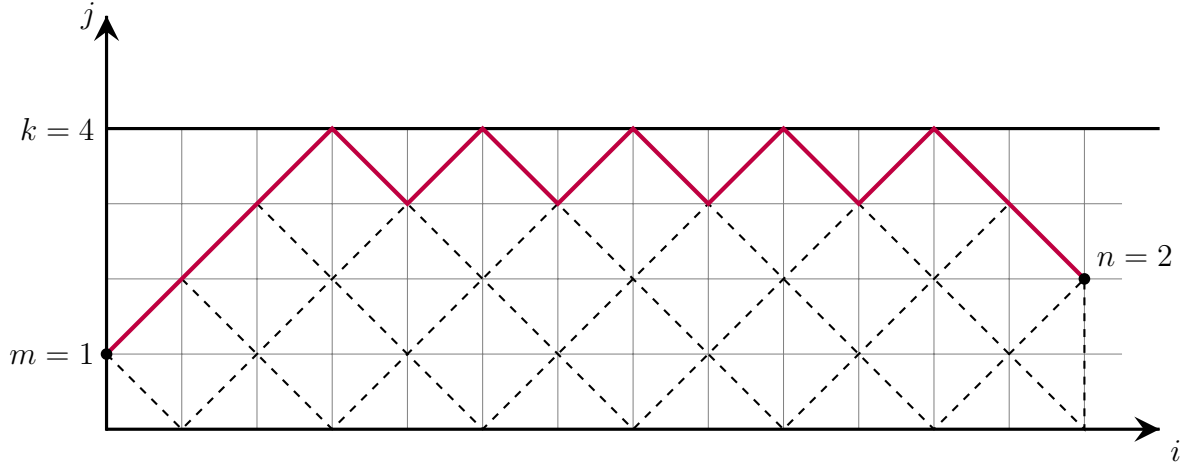
\begin{figure}
\centering 
{
\begin{tikzpicture}[scale=1]

% square grid
\draw[help lines, gray] (0,-0.01) grid (13.5,4.01);

% axes i, j and k
% position: (below/above) + left/right
\tikzset{big arrow/.style={decoration={markings,mark=at position 1 with {\arrow[scale=3,#1,>=stealth]{>}}},postaction={decorate},},big arrow/.default=black}
% Different styles for arrows: >=to (default), stealth, latex
% Alternative type: \tikzset{big arrow/.style={decoration={markings,mark=at position 1 with {\arrow[scale=2,#1,>=to]{>}}},postaction={decorate},},big arrow/.default=black}

\draw[very thick,-] (0,0) -- (14,0);
\draw[very thick,-] (0,-0.02) -- (0,5.5);
\draw[very thick,-] (0,4) node[left] {$k=4$} -- (14,4);

\draw[big arrow] (0,0) -- (14,0) node[below right] {$i$};

\draw[big arrow] (0,-0.02) -- (0,5.5) node[left] {$j$};

% fill and dashed line
% fill=white, yellow, red, etc.

%\draw[ultra thick,dashed,fill=yellow,fill opacity=0.6](0,1)--(1,0)--(2,1)--(1,2)--(0,1);
%\draw[ultra thick,dashed,fill=yellow,fill opacity=0.6](4,1)--(6,3)--(7,2)--(8,3)--(9,2)--(11,4)--(13,2)--%(13,0)--(12,1)--(11,0)--(10,1)--(9,0)--(8,1)--(7,0)--(6,1)--(5,0)--(4,1);

%
\draw[thick,dashed,fill=white,fill opacity=0](0,1)--(1,0)--(4,3)--(7,0)--(10,3)--(13,0)--(13,2);
\draw[thick,dashed,fill=white,fill opacity=0](1,2)--(3,0)--(6,3)--(9,0)--(12,3);
\draw[thick,dashed,fill=white,fill opacity=0](2,3)--(5,0)--(8,3)--(11,0)--(13,2);
%\draw[ultra thick,dashed,fill=white,fill opacity=0](4,1)--(6,3)--(7,2)--(8,3)--(9,2)--(11,4)--(13,2)--(13,0)--(12,1)--(11,0)--(10,1)--(9,0)--(8,1)--(7,0)--(6,1)--(5,0)--(4,1);
%\draw[ultra thick,dashed](5,2)--(6,1)--(7,2)--(8,1)--(9,2)--(10,1)--(11,2)--(12,1)--(13,2);
%\draw[ultra thick,dashed](10,3)--(11,2)--(12,3);

% initial point & terminal point

% path from m=1 to n=2
\draw[ultra thick,purple,-](0,1)node[left] {\color{black} $m=1$}--(3,4)--(4,3)--(5,4)--(6,3)--(7,4)--(8,3)--(9,4)--(10,3)--(11,4)--(13,2)node[above right] {\color{black} $n=2$};

\fill (0,1) circle (2.2pt);
\fill (13,2) circle (2.2pt);

\end{tikzpicture}
}
\caption{\small{A `tiled roof' path of maximal area for $k=4$, starting at $m=1$ and ending at $n=2$, of length $l=2a+n-m=13$,
so for $a=6$, zigzagging $a-k+n = 4$ times at the ceiling. The total area in diamonds is given by the second term in
(\ref{degreek}) and the $\ln q$ coefficient in (\ref{lnGk})
as $(k-n-1)(2a-k+n)/2+a n + (n-m)(n+m-1)/4 = 17.5$, agreeing with the number of diamonds in the figure.}}\label{Tiled roof}
\end{figure}}

We conclude by noting that for $m=n=0$ (excursions) the expression for $G_{k,00} = G_{k}$ simplifies to
\be
\ln G_k (z,q) 
= \sum_{a=1}^{\infty} z^{a} \hskip -0.3cm \sum_{l_1, l_2, \ldots, l_{j} ;\, j\le k\atop { \rm compositions}\;{\rm of}\;a} 
\hskip -0.4cm c_2 (l_1,l_2,\ldots,l_{j} ) ~q^{\sum_{i=1}^j (i-1) l_i}
\ee
an expression identical to the one  for unrestricted walks (\ref{lnG}) but now with compositions restricted to those with
up to $k$ terms.

\section{Keeping track of touchdowns}

A generalization of the generating function considered in the literature is one that also keeps track of the times a path
returns to the floor $n=0$ (a `touchdown'). E.g., the path of fig.\ref{Dyck path} has one touchdown.
Weighing each such touchdown with a factor of $t$, the generating function for paths of length $l$ and area $A$ with $s$
touchdowns becomes
\be
\tG_{k,mn} (t,\z,\q) = \sum_{A,l,s=0}^\infty t^s \, \z^l \, \q^A N_{k,mn;l,A,s}
\ee
(we temporarily reverted to $\z$ and $\q$). Clearly $\tG_{k,mn} (1,\z,\q) = G_{k,mn} (\z,\q)$.
 
This extension can be naturally and effortlessly implemented in our Hamiltonian framework. (In fact, we can equally
effortlessly incorporate arbitrary `markers' to any number of levels, registering the number of passages of the walk
through each of these levels, but we will not pursue this.)
Consider the Hamiltonian $\tH_k$
\be
{\tH}_k = 
\begin{pmatrix}
0\;\;& t\;\, & 0\;\; & 0\;\; \cdots\;\; & 0\;\; & 0\; \\
1 \;\;& 0 \;\;&\q\;\;& 0\;\; \cdots \;\;& 0 \;\;& 0\; \\
0 \;\;& \q\;\; & 0\;\; &\q^{2} \;\cdots\;\; & 0 \;\;& 0\; \\
\vdots \;\;& \vdots \;\;& \vdots \;\;& \;\;\ddots & \vdots\;\; & \vdots\; \\
0\;\; & 0 \;\;& 0 \;\; &0\;\;  \cdots& 0\;\; & \q^{k-1} \;\\
0\;\;& 0 \;\;& 0 \;\; &0 \;\;\cdots & \q^{k-1}\;\;& 0\; \\
\end{pmatrix}
\label{hamtk}\ee
$\tH_k$ is the same as $H_k$ with the exception of the $\ket{0}\bb\bra{1}$ element which is multiplied by $t$.
It should be obvious that $\tH_k^l$ counts area-weighted paths of length $l$, as before, but also multiplies by a factor
of $t$ for each touchdown. Therefore,
\be
\tG_{k,nm} (t,\z,\q) = \sum_{l=0}^\infty \z^l \bra{m} \tH_k^l \ket{n} = \bra{m} (1- \z \tH_k )^{-1} \ket{n}
\ee
as before. Denoting
\be
\tF_k (t,\z,\q) = \det (1-\z \tH_k )
\ee
a calculation entirely analogous to the $t=1$ case yields
\be
\tG_{k,mn} (t,\z,\q) = \z^{n-m}\,  \q^{(n - m)(n+m-1) \over 2} \, {\tF_{m-1} (t,\z,\q) \, F_{k-n-1} (\z \q^{n+1} ,\q) \over \tF_k (t,\z,\q)}
~,~~ n\ge m
\label{Typara}\ee
An expansion of $\tF_k$ in terms of its top row gives
\be
\tF_k (t,\z,\q) = F_{k-1} (\z \q ,\q ) - t\, \z^2 F_{k-2} (\z \q^2 , \q)
\label{tex}\ee
which, together with (\ref{ex}), implies
\be
\tF_k (t,\z,\q) = t F_k (\z,\q) + (1-t) F_{k-1} (\z\q,\q)
\label{trec}\ee
Equations (\ref{Typara}) and (\ref{trec}) are equally valid in terms of variables $z =\z^2$ and $q = \q^2$.
The known expressions for $F_k (z,q)$ allow for the full determination of $\tG_{k,mn} (t,z,q)$.
Direct use of (\ref{Typara},{\ref{trec}) leads to the relation
%\be
%\tG_k (t,z,q) = 1 + t\, z^2 G_{k-1} (zq,q) \tG_k (t,z,q)
%\ee
%and
\be
\tG_{k,mn} (t,z,q) = G_{k,mn} (z,q) \, {t+(1-t) G_{m-1} (z,q) \over t+(1-t) G_k (z,q)}
\label{tGk00}\ee
fully solving the problem. For $m=n=0$, in particular,
\be
\tG_{k} (t,z,q) = {G_{k} (z,q) \over t+(1-t) G_k (z,q)}
\label{hag}\ee
Finally, if we wish {\it not} to count the final touchdown at the end of the path, the corresponding generating
function $\ttG_k (t,z,q)$ becomes
\be
\ttG_k (t,z,q) = 1 + {1\over t} \left( \tG_k (t,z,q) -1 \right)
\ee
and from (\ref{hag})
\be
\ttG_k (t,z,q) = 1+ {G_k (z,q) -1 \over t +(1-t) G_k (z,q)}
\ee

\section{Conclusions}

We presented an intuitive and efficient Hamiltonian framework for representing height-restricted Dyck paths
and evaluating generating functions. These paths were shown to be intimately related to quantum statistics
of exclusion $g=2$, a connection that allowed for the derivation of alternative expressions for the logarithm
of the generating functions.

The use of $\z,\q$ in the determinant calculation (Section 3) and of $z,q$ in the exclusion statistics part
(Section 4), related by (\ref{specc}), may appear somewhat unjustified. We stress, however, that it was simply
done for the convenience of not
carrying $\sqrt z$ and $\sqrt q$ in the initial determinant calculation. In fact, the basic building block $F_k (z,q)$
can be expressed as a determinant of an alternative matrix
\be
F_k (z,q) = \det {\tilde D}_k (z,q) ~, ~~~ {\tilde D}_k (z,q) =
\begin{pmatrix}
1& -1  & 0 & 0\cdots & \bb\bb\bb\bb 0 & \bb\bb\bb 0 \\
-z & 1 &\bb -1 & 0\cdots & \bb\bb\bb\bb 0 & \bb\bb\bb 0 \\
0 & -z q & 1 &-1 \cdots & \bb\bb\bb\bb 0 & \bb \bb\bb 0 \\
\vdots & \vdots & \vdots & \;\;\; \ddots &\bb\bb\bb\bb \vdots &\bb \bb \bb\bb\vdots \\
0 & 0 & 0 & 0 \cdots & \bb\bb\bb\bb 1 &\bb\bb\bb\bb\bb -1 \\
0& 0 & 0  &0 \cdots &\bb\bb\bb\bb -z q^{k-1} & \bb 1 \\
\end{pmatrix}
\ee
involving no square roots. This is guaranteed by the fact that ${\tilde D}_k (z,q)$ above is a similarity
transformation of ${D}_k (z,q)$ in (\ref{Dk})
\be
{\tilde D}_k (z,q) = S \, {D}_k (\z,\q)\, S^{-1}
\ee
with a diagonal matrix $S$ with diagonal elements $\z \q^{n(n-1)/2}$. The entire analysis of section \ref{detsec}
can be done in terms of ${\tilde D}_k$. In fact, using this matrix instead of $D_k$ would also eliminate the
$\z$- and $\q$-dependent prefactor in (\ref{typara}). Using this matrix, however, amounts to counting only
{\it down} steps of the walk and ignoring up steps, assigning them neither length nor area. This is fine in
balanced walks with $m=n$, since for each up step there is an identical down step that accounts for both, but rather
unnatural for unbalanced ones. It also spoils the symmetry $G_{k,mn} = G_{k,nm}$, making the prefactor
reappear (squared) when $n<m$. Altogether, we preferred to keep the more intuitive symmetric Hamiltonian
and secular matrix and change conventions when appropriate.

The recursion relation (\ref{newG}) for $G_k (z,q)$ can be iterated, and with the final condition $G_0 (z,q) = 1$
we obtain\vskip -0.7cm
\be
G_k (z,q) = {1 \over 1- {z \over 1- {z q \over \dots {z q^{k-1} \over 1- z q^k}}}}
\ee
This is a truncated version of the continued fraction related to Ramanujan's identity. The relation of
continued fractions and Dyck walks is known in the mathematics literature, and it is an interesting
topic to investigate the significance of similar truncated fractions for other $G_{k,mn} (z,q)$.

The expressions of length generating functions ($q=1$) for walks with a more general set of up and down steps, and
arbitrary weights associated to each kind of step, have been related to skew-Schur functions \cite{KP}. Including the area
counting variable $q$ would generalize these generating functions to $q$-deformed versions of skew-Schur functions.
This is an interesting topic for further study.

Dyck walks are intimately related to compositions of a large number of $SU(2)$ spins.
In \cite{PS}, the combinatorics and statistics of such compositions were studied using generating function and
partition function techniques. The existence of a ceiling at $n=k$ corresponds to deforming the spin group to
the `quantum group' $SU(2)_{_Q}$ with $Q = \exp[2\pi i/(k+1)]$.
Although the concept of
weighting the paths (spin compositions) with an exponential factor proportional to the area was not relevant there,
which corresponds to taking $q=1$, nevertheless a rich statistical mechanics structure and a corresponding large-$N$ phase transition
were identified. Compositions of spins of arbitrary size $s$, as well as symmetric and antisymmetric spin compositions
were also studied in \cite{PS}, leading to novel statistical properties. The connection and interplay between the two
systems and approaches is an interesting topic that deserves further exploration.

Finally, there are several generalizations of Dyck and Lukasiewicz paths that have been studied in the literature,
and some that have not, which can be treated in a natural way in the Hamiltonian framework. The simplest
one would be Motzkin paths, i.e. walks for which in each step the vertical increment 
is $+1$, $-1$ or 0. This would amount to adding a nonzero diagonal to $H_k$, a relatively benign modification that
can be treated within the present formalism. Generalizations involving larger increments would
add further off-diagonals to $H_k$. Remarkably, these generalizations are also related to quantum exclusion statistics,
but for exclusion higher than 2 and more general one-body spectra. The statistical mechanics of general-$g$ exclusion
systems with an arbitrary discrete energy spectrum have recently been derived using techniques closely related to the
ones in the present work \cite{OP}. Using these techniques the generating functions and statistics of generalized paths
could be studied. We defer treating these cases to a future publication.

\vskip 0.3cm

\noindent
{\bf{Acknowledgments}}

\noindent
We thank Thomas Prellberg for comments on our manuscript and for pointing out the possible
connection to $q$-deformed skew-Schur functions, and the anonymous referees for useful suggestions.
A.P. acknowledges the hospitality of LPTMS at Universit\'e Paris-Saclay,
where this work was initiated.

\end{document}